\documentclass[conference, a4paper]{IEEEtran}

\IEEEoverridecommandlockouts                            
\usepackage{units}
\usepackage{graphicx}
\usepackage{multirow}
\usepackage{amsmath,amssymb,latexsym}
\usepackage{rotating}
\usepackage{lettrine}
\usepackage{textcomp}

\usepackage{hyperref}

\usepackage{xcolor}

\usepackage{lipsum}
\usepackage[pscoord]{eso-pic}
\newcommand{\placetextbox}[3]{
  \setbox0=\hbox{#3}
  \AddToShipoutPictureFG*{
    \put(\LenToUnit{#1\paperwidth},\LenToUnit{#2\paperheight}){\vtop{{\null}\makebox[0pt][c]{#3}}}%
  }%
}%

\title{Predicting Wireless Channel Quality by means of Moving Averages and Regression Models
\thanks{This work was partially supported by the European Union under the Italian National Recovery and Resilience Plan (NRRP) of NextGenerationEU, partnership on ``Telecommunications of the Future'' (PE00000001 - program ``RESTART'').}
}

\author{
    \IEEEauthorblockN{Gabriele Formis\IEEEauthorrefmark{1}\IEEEauthorrefmark{2}\href{https://orcid.org/0000-0001-9290-002X}{\includegraphics[scale=0.65]{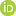}}, Stefano Scanzio\IEEEauthorrefmark{1}\href{https://orcid.org/0000-0001-7643-2342}{\includegraphics[scale=0.65]{orcid_16x16.png}}, Gianluca Cena\IEEEauthorrefmark{1}\href{https://orcid.org/0000-0003-0084-5321}{\includegraphics[scale=0.65]{orcid_16x16.png}}, and Adriano Valenzano\IEEEauthorrefmark{1}\href{https://orcid.org/0000-0003-1238-0808}{\includegraphics[scale=0.65]{orcid_16x16.png}}}
    \IEEEauthorblockA{\IEEEauthorrefmark{1}National Research Council of Italy (CNR--IEIIT), Italy.\\ Email: gabriele.formis@ieiit.cnr.it, stefano.scanzio@cnr.it, gianluca.cena@cnr.it, adriano.valenzano@cnr.it}
    \IEEEauthorblockA{\IEEEauthorrefmark{2}Politecnico di Torino, Italy  \vspace{0.5cm}}
}

\begin{document}
\placetextbox{0.5}{1}{This is the author's version of an article accepted to the IEEE WFCS 2023 conference.}

\placetextbox{0.5}{0.985}{Changes were made to this version by the publisher prior to publication.}
\placetextbox{0.5}{0.97}{The final version of record is available at \href{https://doi.org/10.1109/WFCS57264.2023.10144122}{https://doi.org/10.1109/WFCS57264.2023.10144122}}%
\placetextbox{0.5}{0.05}{Copyright (c) 2023 IEEE. Personal use is permitted.}
\placetextbox{0.5}{0.035}{For any other purposes, permission must be obtained from the IEEE by emailing pubs-permissions@ieee.org.}%

\maketitle
\thispagestyle{empty}
\pagestyle{empty}

\begin{abstract}
The ability to reliably predict the future quality of a wireless channel, as seen by the media access control layer,
is a key enabler to improve performance of future industrial networks that do not rely on wires. 
Knowing in advance how much channel behavior may change can speed up procedures
for adaptively selecting the best channel, making the network more deterministic, reliable, and less energy-hungry, possibly improving device roaming capabilities at the same time.

To this aim, popular approaches based on moving averages and regression were compared, using multiple key performance indicators, on data captured from a real Wi-Fi setup. 
Moreover, a simple technique based on a linear combination of outcomes from different techniques was presented and analyzed, to further reduce the prediction error, and some considerations about lower bounds on achievable errors have been reported.
We found that the best model is the exponential moving average, which managed to predict the frame delivery ratio with a 2.10\% average error and,
at the same time, has lower computational complexity and memory consumption than the other models we analyzed.
\end{abstract}


\section{Introduction}
\label{sec:introduction}
One of the most popular research goals about industrial communications in the past decade has been the use of wireless technologies for connecting machinery and mobile devices over the air \cite{9714203}. 
Industrial networks are demanded to adhere to recent trends like Industry 4.0 \cite{CANAS2021107379}, Industry 5.0 \cite{MADDIKUNTA2022100257}, and the Industrial Internet of Things (IIoT) \cite{8401919}, where the convergence of physical, digital, and virtual environments, and the increased requirements in terms of mobility, sustainability, resilience, human interaction, automatic configurability \cite{9779183}, and (distributed) intelligence pose significant challenges. 

To comply to the new production paradigms and satisfy the previously listed goals, industrial networks are becoming increasingly heterogeneous \cite{SCANZIO2021103388}, which implies that the requirements they are expected to meet in terms of reliability, determinism, and timeliness are becoming more and more strict. 
Research works aimed at improving the different wireless communication technologies to meet these constraints 
\cite{9945847, electronics11030304, LEONARDI202257, 2016-TII-WiRed, 9921559}
represent concrete contributions toward their widespread adoption in the industry. 
In addition, research on (ultra-)low power solutions and energy-saving mechanisms are of utmost importance 
for those technologies employed in mobile and battery-powered devices \cite{9903301}.

The quality of the wireless channel, and particularly its reliability in terms of the frame delivery ratio (FDR), may change quickly and unpredictably, due for instance to the interference and disturbance from the surroundings or the mobility of communication devices and objects within the considered environment \cite{5502548}. 
A fair amount of effort has been spent in the past years to characterize communication over wireless channels \cite{705532}. Many research studies aim at evaluating the probability that a symbol (or an entire frame) is affected by an error starting from the physical characteristics of the environment (distance between nodes, signal reflections, electromagnetic noise, bit rate, and so on). These approaches are mostly useful in the design phase of distributed systems where nodes are interconnected over the air, as they permit to estimate what communication quality will be experienced by them, and hence to properly configure both network and application parameters.

The ability to measure the quality of communication once a system has been deployed can be useful to continuously tune such parameters, in such a way to optimize data transmission and try to ensure some properties for it. 
For instance, the Minstrel algorithm monitors the outcomes of transmission attempts in commercial Wi-Fi drivers and adaptively selects the best modulation and coding scheme (MCS) to maximize the chance of success. 
In the case some messages are characterized by firm deadlines that must not be exceeded (or, at least, the likelihood that they are missed must stay below some given threshold), the ability to foresee the behavior of the wireless link in the near future could drive some mechanisms that proactively stop background transmission of non-critical data when communication quality is expected to worsen.
More in general, if the future quality of the channel could be inferred from the outcomes of past frame exchanges, applications were enabled to use this information to proactively counteract channel quality deterioration by, e.g., switching the communication channel to a better quality one, decreasing the amount of best-effort traffic to prioritize real-time traffic, varying some parameters related to the communication protocol, or activating some redundant communication path.

Applying machine learning, and in particular artificial neural networks (ANN), to Wi-Fi \cite{9786784}, was exploited in \cite{9120030} on artificial data, in \cite{8884240} to predict the received signal strength, and in \cite{8813020, 9781119562306} to predict the channel gain in specific applications. 
In \cite{2022-ITL-ML, 9921698} it was instead applied to a database acquired on real devices deployed in a real environment, hence showing that the future quality of a wireless channel, expressed in FDR terms, can be predicted with an error that is deemed acceptable for many application contexts.

To make results comparable, this work relies on the very same database employed in \cite{2022-ITL-ML, 9921698}.
However, we now analyze a number of known algorithms that require very simple parameterization if compared to ANNs (in which a huge amount of data is needed to estimate the model parameters, i.e., weights and biases). 
In particular, the models we consider here are based on moving averages and regression: simple moving average (SMA), weighted moving average (WMA), exponential moving average (EMA), simple linear regression (SLR), and polynomial regression (PR) of degrees two and three. All these methods, with the exception of EMA, are completely characterized by the number of samples considered in the past, which is the only hyperparameter that needs to be optimized in the training process.

An extensive experimental campaign was performed to compare these models individually and with respect to the ANN. 
Moreover, we also considered heuristics based on linear combinations of the FDR estimates provided by the above models, to improve prediction accuracy further.
Results show that the prediction error of EMA is comparable with ANN, but with a dramatically lower computational complexity and memory consumption.

The remaining part of the paper has the following structure: in Section~\ref{sec:channel_quality} the problem of predicting the channel quality in terms of the frame delivery probability is introduced, while in Section~\ref{sec:models} the models based on regression and moving average are described and their applicability to the problem analyzed here is discussed.
The experimental setup is presented in Section~\ref{sec:setup}, experimental results are exposed in Section~\ref{sec:results}, and finally Section~\ref{sec:conclusions} concludes this work.

\section{Channel quality prediction}
\label{sec:channel_quality}
This work is specifically based on \mbox{Wi-Fi}, but the proposed methods can be applied with non-substantial changes also to other communication technologies. 
To monitor the quality of the channel in FDR terms (i.e., the fraction of correctly delivered frames over the total number of frames transmitted on air), a cyclic probing was employed by performing confirmed one-shot data frame transmissions (i.e., by setting the retry counter to $0$, which disables retransmission)
between two nodes with period $T_s=\unit[0.5]{s}$.
Backoff was also disabled, by setting the contention window to $0$,
which implies that every transmission attempt may be deferred at worst once by an ongoing transmission of another wireless device operating in the same frequency band (interference).
As a consequence, channel quality is sampled at $\unit[2]{Hz}$ and samples are evenly spaced over time.

The outcome of the transmission of the $k$-th frame
(or, equivalently, of the $k$-th channel probing attempt)
can be either \textit{success} ($x_k=1$) when the related Acknowledgement (ACK) frame comes back to the sender, or \textit{failure} ($x_k=0$) in the case the ACK frame is not received within the related timeout.
Failure events can be related to either the loss of the data frame or the correct reception of the data frame and the subsequent loss of the ACK. 
This means that the quality of the channel is analyzed from the point of view of the sender.
This represents what is typically of interest in real applications: the sender node, which is responsible for the retransmission of those frames that went lost (identified by the lack of the ACK), is also the entity that can exploit predictions about channel quality to implement suitable countermeasures.

\begin{figure}[t]
	\begin{center}
	\includegraphics[width=0.95\columnwidth]{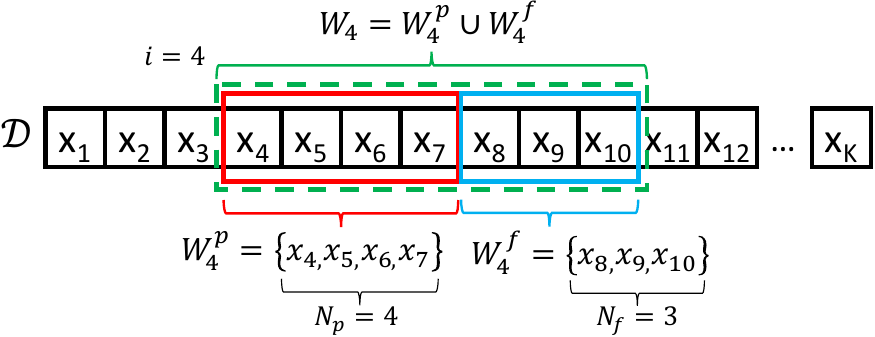}
	\end{center}
	\caption{Management of the moving window $W_i$ (past and future).}
	\label{fig:sliding}
\end{figure}

Let the sequence of outcomes $\mathcal{D}=\{x_1,...,x_k,...,x_K\}$ be the database used in this analysis, where $K=|\mathcal{D}|$ is its size.
As customary in machine learning, two separate sample sets were defined: the \textit{training} database $\mathcal{D}_{tr}$ used to train the model (if needed), and the \textit{test} database $\mathcal{D}_{te}$ used to check the quality of the model. 
A moving window of width $N_p+N_f$ is defined, where $N_p$ is the number of samples used to compute the prediction and $N_f$ the number of future samples on which the target (i.e., the quantity to be predicted) is evaluated. 
Given a database with $K$ elements, and depending on the starting position,  
a number of distinct windows $W_i$ exist, where $i=1,...,K-N_p-N_f+1$.
Each such window can be expressed as $W_i = W^p_i \cup W^f_i$ where $W^p_i=\{x_j| i \leq j \leq i+N_p-1\}$ are the $N_p$ samples used to perform the prediction and $W^f_i=\{x_j|i+N_p \leq j \leq i+N_p+N_f-1\}$ are the $N_f$ future samples exploited for calculating the target $t_i$ (see Fig~\ref{fig:sliding}). 
The duration of these two windows is $T^p=N_p \cdot T_s$ and $T^f = N_f \cdot T_s$, respectively.

The target $t_i$ is computed as the arithmetic mean of the samples included in $W^f_i$
\begin{equation}
t_i = \frac{1}{N_f}\sum_{j=i+N_p}^{i+N_p+N_f-1} x_{j},
\label{eq:target}
\end{equation}
and is a statistical estimation, evaluated on a limited number of samples,
of the frame delivery probability $\epsilon_i$ observed in the future window.

The fact that the target $t_i$ is just an estimate of $\epsilon_i$, and not the quantity $\epsilon_i$ we are looking for,
makes the kind of problem analyzed in this work more complex than typical time series analyses.
The goal of this work is to find a \textit{prediction function} $f(\cdot)$ that, given $W^p_i$,
evaluates a satisfactory estimate $y_i$ of the target $t_i$ calculated over $W^f_i$,
\begin{equation}
  y_i = f(W^p_i).
\label{eq:pred}
\end{equation}

Suitable quantities can be used to evaluate the accuracy of the prediction obtained from (\ref{eq:pred}).
In particular, the absolute and squared errors were taken into account
\begin{eqnarray}
e_i & = & |t_i-y_i|,\\
e_i^2 & = & (t_i-y_i)^2,
\end{eqnarray}
where the $|\cdot|$ operator represents the absolute value. 
Starting from these errors, the corresponding mean absolute error (MAE) and mean squared error (MSE) can be obtained by averaging $e_i$ and $e_i^2$ for all 
windows in a given test database
\begin{eqnarray}
\operatorname{MAE} & = & \frac{1}{N_{te}} \sum_{i=1}^{N_{te}} e_i,\\
\operatorname{MSE} & = & \frac{1}{N_{te}} \sum_{i=1}^{N_{te}} e_i^2,
\end{eqnarray}
where 
$N_{te} = |\mathcal{D}_{te}| - N_p - N_f +1$. 

\begin{figure}[t]
	\begin{center}
	\includegraphics[width=0.95\columnwidth]{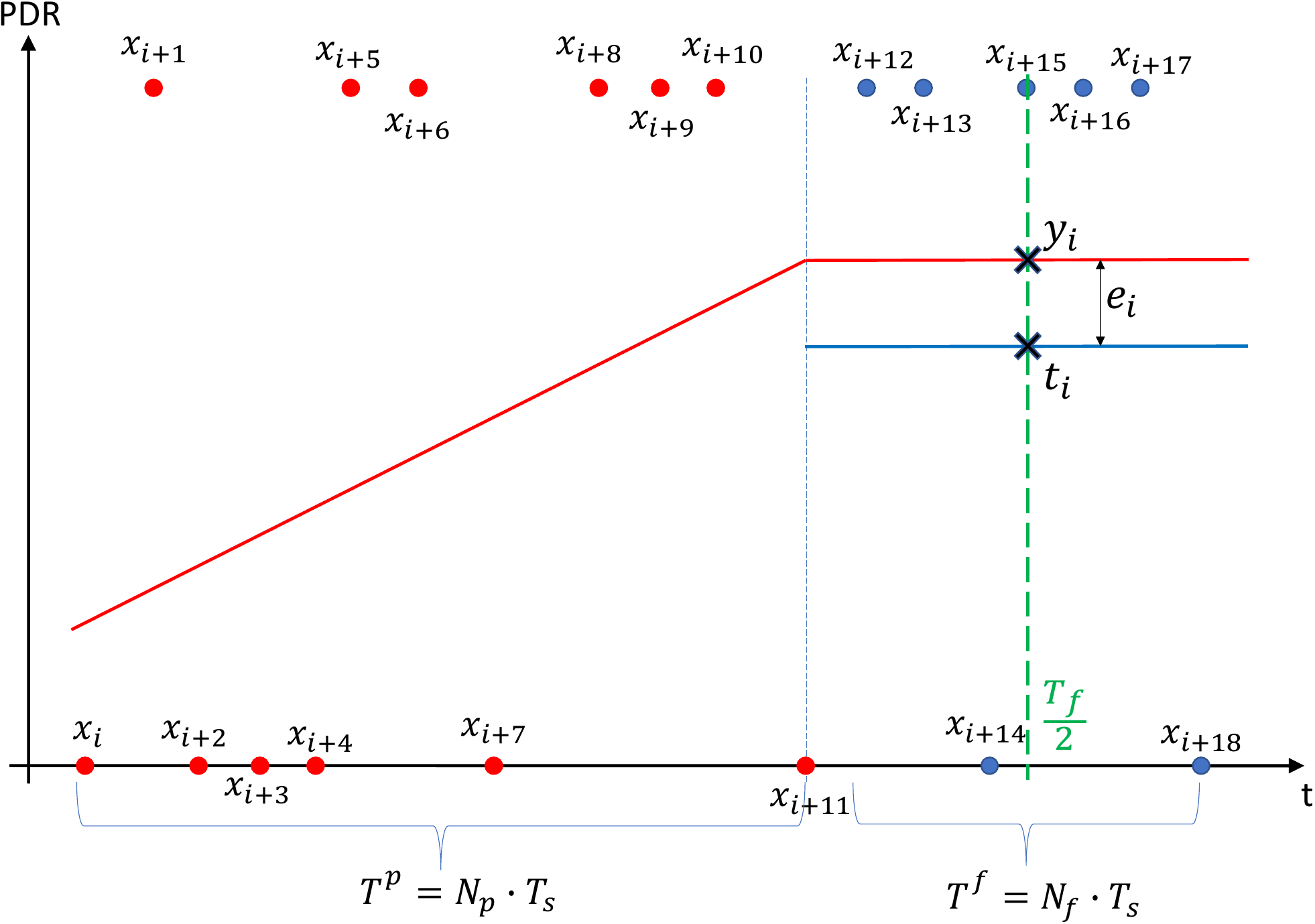}
	\end{center}
	\caption{Example of application of SMA for predicting the future quality of the Wi-Fi channel ($N_p=12$, $N_f=7$).}
	\label{fig:SMA}
\end{figure}

In Fig.~\ref{fig:SMA}, the prediction process is exemplified by using SLR as the prediction function. 
The $N_p$ samples (outcomes) reported in the left part of the plot are used for estimating the two parameters of the interpolating line $y=\beta_1 \cdot t + \beta_0$, i.e., slope $\beta_1$ and intercept $\beta_0$. 
It can be formally demonstrated that applying SLR to Boolean values $x_i$ yields unbiased estimators for both $\beta_1$ and $\beta_0$.
The proof was not included for space reasons.
After computing the two model parameters, 
an estimate for FDR is obtained by evaluating the prediction function in the most recent sample of the interval $W^p_i$, that is, $x^*_i = x_{i+N_p-1}$ ($x_{i+11}$ in the example of Fig.~\ref{fig:SMA}). 
Then, it is assumed that the FDR in the future window $W^f_i$ remains at the same level as the above estimate, i.e., $y_i=\beta_1 \cdot x^*_i + \beta_0$ (piecewise linear interpolation).

We found that, for the kind of data analyzed in this work, computing the estimate using the true regression model (linear, quadratic, etc.) in the middle point $t^*$ of $W^f_i$ (i.e., $\frac{T^f}{2}$ after its beginning) actually provided lower accuracy. 
In particular, the prediction $y_i=\beta_1 \cdot t^* + \beta_0$ (as well as those related to higher-order regression) is unsatisfactory when $t^*$ is located far from the most recent sample used for estimation. 
This statement has been verified experimentally. 
Results for true, non-piecewise regression have been reported only for SLR, and are identified with the name ``predictive SLR'' (PSLR).

\section{Prediction models}
\label{sec:models}
Besides SMA, which was already checked against an ANN model in \cite{9921698}, in this work we thoroughly analyze the prediction capability of regression models like SLR and PR, as well as other models based on moving averages like WMA and EMA. 
Then, a lower bound on the achievable error has been obtained by selecting the best model on a frame-by-frame basis with an oracle.
In other words, we assume that an oracle is available that tells us which prediction model achieves the highest accuracy at any given time.
Finally, the models have been used together by calculating the estimate as a linear combination of their predictions.

Polynomial regression models were successfully exploited in other application contexts like clock synchronization protocols \cite{MONGELLI20161, 6817598}, as well as the prediction of rainfall \cite{4769479} and movements \cite{8747047}. 
Although the results obtained in those cases were very promising, there is no evidence that the same performance could be obtained also in this application domain.

\begin{table}[t]
  \caption{Prediction models and number of model parameters.}
  \label{tab:models}
  \footnotesize
   
  \begin{center}
    \tabcolsep=0.18cm
    \def\arraystretch{1.28}
    \begin{tabular}{c|l|c}
    Model & Equation & Param. \\
    \hline \hline
    SMA & $y^{\mathrm{SMA}}_i=\beta_0 = (x_{i+N_p-1}+...+x_{i})/N_p$ & $1$ \\
    WMA & $y^{\mathrm{WMA}}_i=w_1 \cdot x_{i+N_p-1}+...+w_{N_p} \cdot x_{i}$ & $1$ \\
    EMA & $y^{\mathrm{EMA}}_i=\alpha \cdot x_{i+N_p-1} + (1-\alpha) \cdot y^{\mathrm{EMA}}_{i-1}$ & $0$\\
    SLR & $y^{\mathrm{SLR}}_i=t \cdot \beta_1 + \beta_0$ & $2$ \\
    PR2 & $y^{\mathrm{PR2}}_i=t^2 \cdot \beta_2 + t \cdot \beta_1 + \beta_0$ & $3$\\
    PR3 & $y^{\mathrm{PR3}}_i=t^3 \cdot \beta_3 + t^2 \cdot \beta_2 + t \cdot \beta_1 + \beta_0$ & $4$\\
    \hline
   \end{tabular}
  \end{center}
\end{table}

In the following we distinguished between hyperparameters, which are optimal values determined in the training phase ($N_p$ for SMA, WMA, and regression, $\alpha$ for EMA), and parameters specific of every model, which are dynamically evaluated in the test phase
using different techniques (moving averages, interpolation, etc.).
Table~\ref{tab:models} summarizes the prediction models we analyzed,
along with the number of their specific parameters. 
SMA, WMA, and EMA, rely on a single model parameter,
whereas two are required for SLR (slope $\beta_1$ and intercept $\beta_0$ of the regression line). 
For second-order polynomial regression (PR2) the number of parameters is three (i.e., $\beta_2$, $\beta_1$, and $\beta_0$), and four for third-order polynomial regression (PR3).

Parameter $\beta_0$ for SMA (which can be seen as both a moving average and a zero-degree regression) is obtained by averaging the most recent $N_p$ samples. 
For WMA, the average is weighted using $w_j$ coefficients with $j=1,...,N_p$ (see Table~\ref{tab:models}), where the coefficient associated with the most recent sample $x_{i+N_p-1}$ is $w_1=\frac{N_p}{N_p \cdot (N_p+1)/2}$, the one assigned to an intermediate sample is $w_j=\frac{N_p-j+1}{N_p \cdot (N_p+1)/2}$, and that assigned to the oldest sample $x_{i}$ is $w_{N_p}=\frac{1}{N_p \cdot (N_p+1)/2}$. 
For instance, if $N_p=3$ the three weights are $w_1=\frac{3}{6}$, $w_2=\frac{2}{6}$, and $w_3=\frac{1}{6}$.

Regarding EMA, the new prediction $y^{\mathrm{EMA}}_i$ is obtained as a linear combination of the most recent sample $x_{i+N_p-1}$ and the old prediction $y^{\mathrm{EMA}}_{i-1}$, where $\alpha$ is used to weight the two contributions (see the related formula in Table~\ref{tab:models}). 
Basically, it corresponds to the simplest form of infinite impulse response (IIR) low-pass filter,
which is applied here to the outcomes of the channel probing process.
A value of $\alpha$ near to $1$ gives higher priority to the new sample, while a value of $\alpha$ near to $0$ privileges the old prediction. 
This is the same as considering a WMA model where all samples are weighted in exponentially decreasing way: 
$y^{\mathrm{EMA}}_i = \alpha \cdot x_{i+N_p-1} + \alpha (1-\alpha) \cdot x_{i+N_p-2} + \alpha (1-\alpha)^2 \cdot x_{i+N_p-3} + \alpha (1-\alpha)^3 \cdot x_{i+N_p-4} + \cdots$. 

Finally, the parameters for SLR, PR2, and PR3 are obtained by finding the interpolating polynomial of degree $1$, $2$, and $3$, respectively, that minimizes the squared error of the samples contained in $W^p_i$.
In theory, increasing the regression order (from SLR to PR2 to PR3) may provide models that, in some circumstances, better fit the non-linearity of the channel they have to predict.
In practice, having more parameters increases the likelihood of overfitting.
Moreover, model accuracy worsens because a larger number of parameters have to be estimated with the same number $N_p$ of samples.

All models make use of $N_p$ samples for estimation, with the exception of EMA that only relies on the last sample.
From a computational point of view, EMA requires very simple operations, which makes it the best candidate for devices with limited resources in terms of memory, computational power, and energy.

\subsection{Estimation of Hyperparameters ($N_p$ and $\alpha$)}
\label{sub:estimation}
\begin{figure*}[t]
    \begin{center}
    \includegraphics[width=2.00\columnwidth]{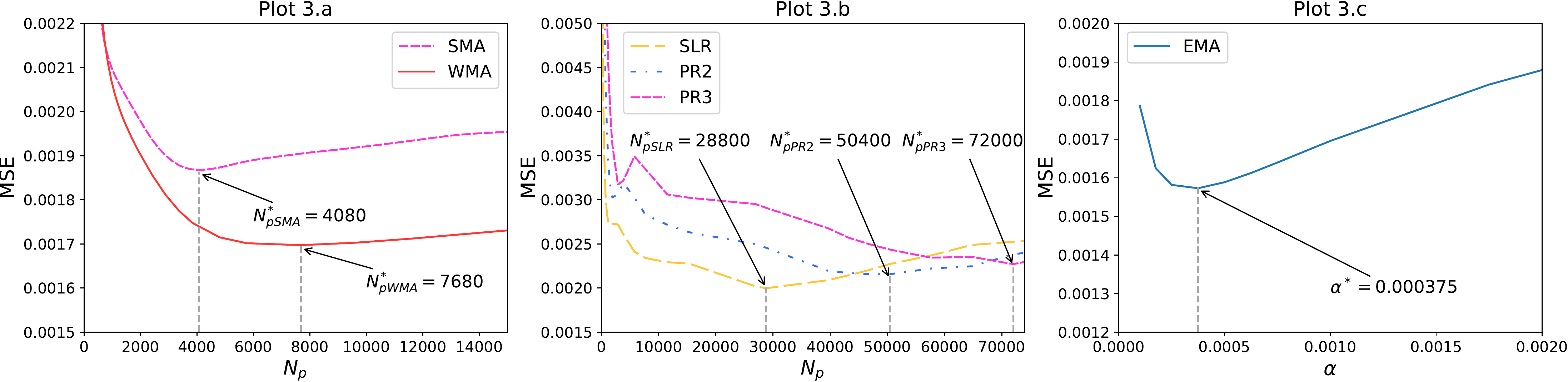}
    \end{center}
    \vspace{-0.1cm}
    \caption{Prediction accuracy (MSE) during training 
    vs. model hyperparameters ($N_p$, $\alpha$) 
 and estimation of their optimal values ($N^*_p$, $\alpha^*$).}
	\label{fig:estimation}
\end{figure*}

The number of samples $N_p$ used to perform the estimation is an important aspect of the model and heavily affects the achievable performance.
The optimal value for it, identified with the symbol $N_p^*$, was evaluated over the training database $\mathcal{D}_{tr}$ by varying the size of the past window and by selecting the one that provides the lowest mean squared error
\begin{equation}
    \label{eq:sma_opt}
    N^*_\mathrm{p} = \arg \min_{N_p} \frac{1}{N_{tr}} \sum_{i = 1}^{N_{tr}}
\Bigl(t_i - f\bigl(W^p_i(N_p)\bigr) \Bigr)^2
\end{equation}
where 
$N_{tr} = |\mathcal{D}_{tr}| - N_p - N_f +1$, 
$N_p^\mathrm{min} \leq N_p \leq N_p^\mathrm{max}$,
while the notation $W^p_i(N_p)$ stresses the fact that function $f(\cdot)$ is evaluated over a moving window that includes $N_p$ samples in the past.

The optimal value $N_p^*$, which is estimated on the training database $\mathcal{D}_{tr}$ using \eqref{eq:sma_opt}, 
was then used with the test database $\mathcal{D}_{te}$ to obtain the relevant statistics. 
Despite $N_p^*$ is expected to achieve the best accuracy in the test too, if the quantity to be predicted does not vary excessively over time a larger $N_p$ value could be selected in order to decrease the sampling error.
Otherwise, in quickly varying spectrum conditions, smaller $N_p$ values permit to more accurately track the evolution of FDR.

For methods like EMA, which do not depend on $N_p$, equation \eqref{eq:sma_opt} cannot be directly applied. Nevertheless, a very similar procedure was used to find the optimal value for the $\alpha$ coefficient (denoted $\alpha^*$) in the training set $\mathcal{D}_{tr}$, which was subsequently used in the test phase to assess prediction results.

\subsection{Oracle}
\label{sub:oracle}
Let $y_i^m$ be the prediction for the FDR in the future window $W^f_i$ obtained by  
model $m \in \mathcal{M}$, where 
$\mathcal{M} = \{ \mathrm{SMA},\mathrm{WMA},\mathrm{EMA},\mathrm{SLR},\mathrm{PR2},\mathrm{PR3} \} $
is the set of basic models we considered in this paper.
The term \textit{oracle predictor} (OR) denotes an additional model that, on a sample-by-sample basis, 
selects the specific prediction function (among those listed above) that provides the lowest squared error
\begin{equation}
\label{eq:OR}
y^{\mathrm{OR}}_i = y_i^{\arg \min_{m \in \mathcal{M}
} \bigl( t_i - y_i^{(m)} \bigr)^2}.
\end{equation}

Unfortunately, there is no algorithm able to single out the best model in advance, 
as always performing the right choice would require $t_i$ to be known.
As a consequence, the error achieved by OR represents sort of a lower bound for the class of mechanisms 
where prediction relies on the selection among a finite number of available options.

\subsection{Linear combination}
The last model proposed in this work,
we term COM,
performs the prediction by combining the outputs of two or more of the above models. 
In particular, the accuracy of a simple linear combination was analyzed and compared with all the other models 
in $\mathcal{M}$
considered individually.
The COM model is described by the equation
\begin{equation}
\label{eq:int_model}
    y^{\mathrm{COM}}_i = \sum_{m \in \mathcal{M}
    } \lambda^{(m)} \cdot y_i^{(m)},
\end{equation}
and is parameterized by 
the tuple $\{ \lambda^{(m)}, m \in \mathcal{M} \}$,
where $\sum_{m \in \mathcal{M}} \lambda^{(m)} =  1$,
which defines the specific weights applied to every basic model $m$ to derive the overall prediction $y^{\mathrm{COM}}_i$.

\section{Experimental setup}
\label{sec:setup}
The database we used for 
evaluating the accuracy of prediction models
was acquired on a real setup, 
deployed in our lab, which includes four TP-Link TL-WDN4800 Wi-Fi adapters installed in two Linux PCs and four access points.
PCs and APs were located $\unit[3 \div 4]{m}$ meters apart, so that signal attenuation was negligible.
Slightly more than $15$ active APs were visible from our testbed,
which caused mild interference, especially on working days.
According to \cite{2022-ITL-ML}, the FDR for the test database $\mathcal{D}_{te}$ lain in the range $60 \div 83\%$, while for the training database $\mathcal{D}_{tr}$ it usually stayed between $55$ and $90\%$, with a single negative peak where the FDR approached $35\%$.

Adapters were configured to transmit with a fixed speed of $\unit[54]{Mb/s}$. 
Automatic retransmission, backoff procedure, and frame aggregation were disabled,
so that channel conditions could
be periodically probed every $T_s=\unit[0.5]{s}$ with packets whose size is $\unit[50]{B}$. 
The quite large value we selected for $T_s$ provides an adequate certainty that every transmission attempt is completely ended 
(i.e., either the ACK frame is received or an ACK timeout event is raised by the adapter) before the next one is issued. 
At the same time, it caused negligible perturbations to the environment.
Such an arrangement is possible thanks to the \texttt{ath9k} driver, which permits the source code of the driver to be modified so as to customize some operating parameters and behaviors of the media access control (MAC) layer.

For collecting and storing the $x_k$ samples, the software-defined MAC (SDMAC) framework \cite{7991945,8477080}  was used, which is a set of rules and the related application programming interface (API) that permit the outcome of frame transmissions to be notified (along with ancillary information) to an application running in user space from a purposely modified driver.
A specific program was developed aimed at cyclically sending data
frames and collecting the transmission outcomes in a database. 
In addition to $x_k$, the information recorded
in the database for every frame includes the sending time, the reception time of the related ACK, and the received signal strength (RSS)
of the received ACK frame, but currently only $x_k$ transmission outcomes are exploited for prediction. 
It is worth remarking again that the events we collect do not correspond to frame arrivals to destination, 
but concern the ACK frames automatically returned by the recipient every time a frame is correctly received. 
Therefore, it may happen that a frame arrives to the recipient correctly but the related ACK is lost, this implying that the recorded outcome is $x_k=0$. 
This behavior is completely coherent with real-world distributed applications, where channel quality measurement is done by the transmitter node, which exploits this information to optimize communication parameters for, e.g., lowering power consumption, increasing reliability, or decreasing latency.

\begin{table*}[t]
  \caption{Comparison among prediction models (SMA, WMA, EMA, SLR, PR2, PR3, PSLR, COM) based on statistics about errors.}
  \label{tab:res_global}
  \small
  \begin{center}
    \tabcolsep=0.22cm
    \def\arraystretch{1.25}
    \begin{tabular}{ccc|c|ccccccc|c}
    Model & Optimal Param. & Used Param. & $\mathrm{MSE}$ & $\mathrm{MAE}$ & $\sigma_e$ & $e_{\mathrm{p}_{90}}$ & $e_{\mathrm{p}_{95}}$ & $e_{\mathrm{p}_{99}}$ & $e_{\mathrm{p}_{99.9}}$ & $e_{\mathrm{max}}$ & $w$ \\
    & &  & $[\cdot 10^{-3}]$ & \multicolumn{7}{c|}{[\%]} & [\%] \\
    \hline
    SMA & $N_p^*=4080$ ($\unit[34]{m}$)  & $N_p=4080$ & 0.987 & 2.34 & 2.10 & 5.04 & 6.40 & 9.77 & 14.49 & 15.38 & 12.57 \\
    WMA & $N_p^*=7680$ ($\unit[\sim 1]{h}$)  & $N_p=7680$ & 0.879 & 2.24 & 1.94 & 4.75 & 5.87 & 8.75 & 13.92 & 14.91 & 7.54 \\
    EMA & $\alpha^*=0.000375$ & $\alpha=0.000375$       & 0.803 & 2.16 & 1.83 & 4.52 & 5.58 & 8.52 & 13.53 & 14.86 & 12.54 \\
    SLR & $N_p^*=28800$ ($\unit[4]{h}$) & $N_p=28800$ & 0.982 & 2.43 & 1.98 & 5.11 & 6.31 & 8.26 & 13.43 & 13.89 & 27.58 \\
    PR2 & $N_p^*=50400$ ($\unit[7]{h}$) & $N_p=28800$ & 1.362 & 2.68 & 2.54 & 5.85 & 7.35 & 11.48 & 18.44 & 19.48 & 19.59 \\
    PR3 & $N_p^*=72000$ ($\unit[10]{h}$) & $N_p=28800$& 1.437 & 2.81 & 2.55 & 6.25 & 7.74 & 11.98 & 18.35 & 19.50 & 20.18 \\ 
    \hline
    PSLR& $N_p^*=28800$ ($\unit[4]{h}$) & $N_p=28800$ & 1.058 & 2.51 & 2.07 & 5.30 & 6.57 & 8.57 & 13.84 & 14.35 & -  \\
    \hline
    COM & - & $N_p=28800$ & 0.920 & 2.28 & 2.01 & 4.92 & 5.98 & 9.35 & 14.28 & 15.46 & - \\
    \hline
    ANN \cite{9921698}* & - & $N_p=14400$ &  0.699 & 2.04 & - & 4.16 & 5.22 & 7.78 & 12.12 & - & - \\
    \hline
    \end{tabular}
    \scriptsize{\\ *Results related to ANN are those reported in \cite{9921698}. They were obtained with a slightly higher number of samples, due to the different $N_p$ values used in that work.}
    \end{center}
\end{table*}

Databases are recorded concurrently for channels $1$ ($\unit[2.412]{GHz}$), 
$5$ ($\unit[2432]{GHz}$), $9$ ($\unit[2.452]{GHz}$), and $13$ ($\unit[2.472]{GHz}$). 
All the databases used in this paper refer to channel $13$. 
In particular, the training database $\mathcal{D}_{tr}$ consists of $2,807,524$ samples 
(more than $16$ days), while the test database includes $460,927$ samples (more than $2.5$ days).

\section{Results}
\label{sec:results}

The procedure described in Subsection~\ref{sub:estimation} was applied to the training database $\mathcal{D}_{tr}$ to obtain the optimal configuration for all hyperparameters. 
For the computation of $t_i$, $N_f=3600$ samples
were included in the future window $W^f_i$, which correspond to a duration $T^f=\unit[30]{min}$.
Hence, the estimate of the channel FDR can be thought of as referred to the time $\frac{T^f}{2}=\unit[15]{min}$ relative to the beginning of $W^f_i$. 

The MSE obtained for the SMA and WMA models versus the value $N_p$ is reported in Plot~\ref{fig:estimation}.a. 
Dashed vertical lines highlight the minimum of these curves, and represent the value $N^*_p$ for any specific model. 
Plot~\ref{fig:estimation}.b (in the middle) has the same meaning as Plot~\ref{fig:estimation}.a, the only difference being that it refers to the SLR, PR2, and PR3 models. 
They were reported separately because of the very different ranges for $N_p$. 
Finally, in the rightmost Plot~\ref{fig:estimation}.c related to EMA, the hyperparameter exploited for minimizing the MSE is $\alpha$.
The optimal value is denoted $\alpha^*$.
The optimal hyperparameter for every model is reported in the second column of Table~\ref{tab:res_global}.

\subsection{Comparison}
The optimal values of $N^*_p$, which minimize the MSE for every model ($\alpha^*$ for EMA), have been used for the comparison in the test phase.
We found that $N^*_p$ is equal to $50400$ for PR2 and $72000$ for PR3, which correspond to $\unit[7]{h}$ and $\unit[10]{h}$, respectively. 
Likely, using for the estimation a number of past samples that span over a period longer than $\unit[4]{h}$ does not suit most application contexts. 
As a consequence, for the PR2 and PR3 models we decided to report results obtained by setting $N_p=28800$, which corresponds to $\unit[4]{h}$. 
The hyperparameters actually used for the experimental evaluation are reported in the third column of Table~\ref{tab:res_global}.

In the same table, comparative results among the presented models have been reported. 
Concerning the squared error we only show $\mathrm{MSE}=\mu_{e^2}$, 
while for the absolute error several statistical indexes have been reported because they are more meaningful from an intuitive point of view. 
In particular, in addition to $\mathrm{MAE}=\mu_{e}$, the reported statistics include the \textit{standard deviation} ($\sigma_e$), \textit{percentiles} ($e_{\mathrm{p}_{90}}$, $e_{\mathrm{p}_{95}}$, $e_{\mathrm{p}_{99}}$, $e_{\mathrm{p}_{99.9}}$), and \textit{maximum} ($e_{\mathrm{max}}$).
For instance, $99\%$ of the times $e_i$ lies below the 99-percentile $e_{\mathrm{p}_{99}}$, while only $1\%$ of the predictions suffer from errors above that value.

Finally, the last column $w$ of the table represents the percentage by which any given method wins over the others. 
As an example, $w=12.57\%$ for SMA means that the $12.57\%$ of the times the SMA model suffered from an absolute error $e_i$ (or, equivalently, $e_i^2$) that is lower than all the other models.

By observing the obtained results it can be seen that the EMA model provided tangibly better accuracy than the others, especially for what concerns MAE, $\sigma_e$, and low-order percentiles.
Showing the superiority of EMA is one of the most important outcomes of this work, 
since it is the model with the lower requirements in terms of computational power and memory, and consequently energy consumption. 
This makes it suitable to perform channel quality prediction also in small embedded devices with limited resources.

\begin{figure}[t]
	\begin{center}
	\includegraphics[width=0.95\columnwidth]{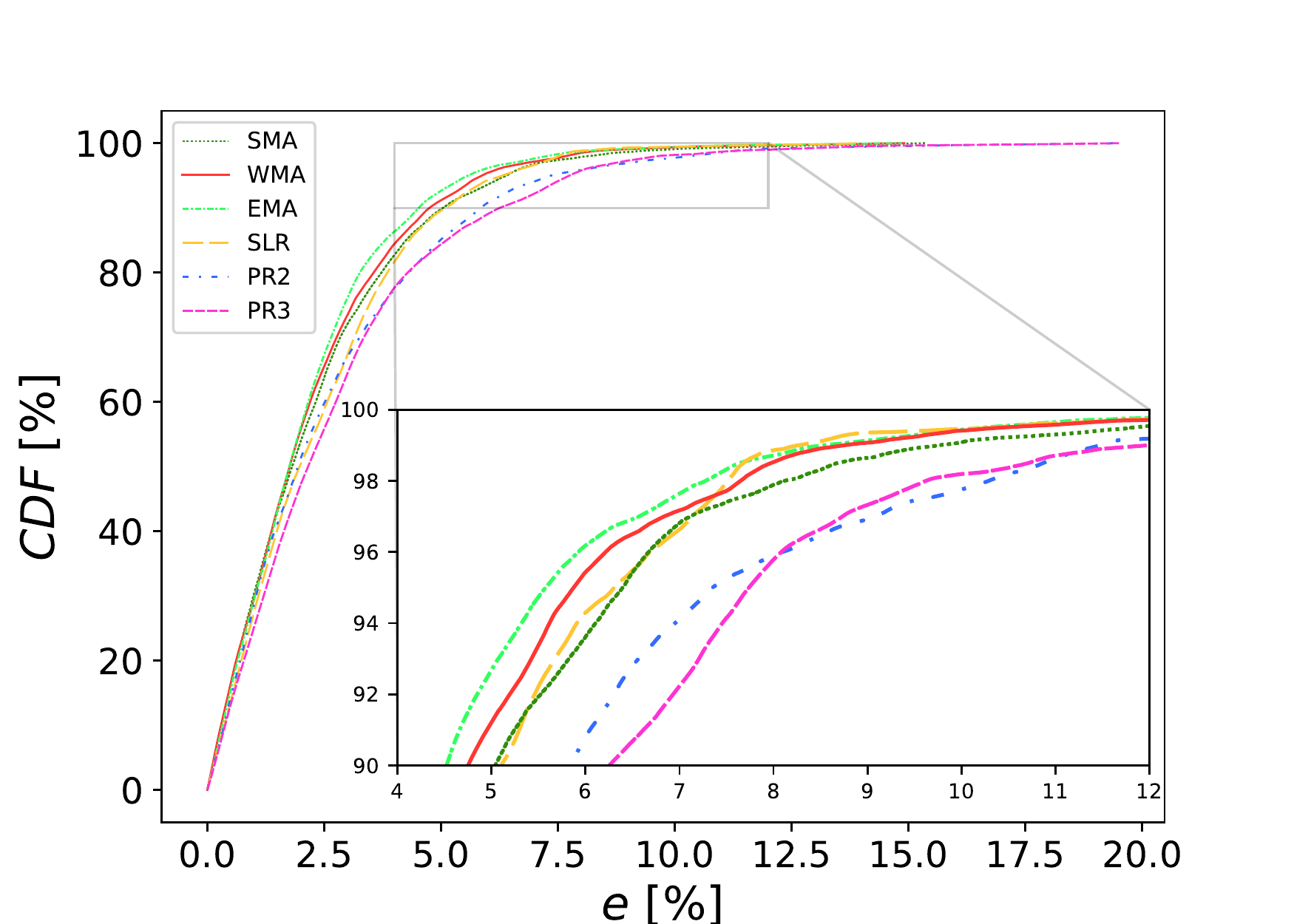}
	\end{center}
	\caption{CDF of the prediction error for the analyzed models.}
	\label{fig:CDF}
\end{figure}

The two models based on moving average, WMA and SMA, provided sensibly worse performance when compared to EMA.
As expected, WMA experienced lower errors than SMA due to its ability to track channel variations by focusing more on recent samples.

Models based on regression such as SLR, PR2, and PR3 showed higher errors, which worsen with the regression order. 
As previously pointed out, the direct use of regression models to infer the behavior of the channel in a future instant in time, as for PSLR, was always pejorative. 
As an example, the MAE of PSLR ($2.51\%$) is greater than SLR ($2.43\%$). 

The COM model with $\lambda^{(m)}=\frac{1}{6}$ (all the six considered models are equally weighted) did not provide any advantages in terms of the prediction accuracy. 
Finally, the last row of the table reports the experimental results obtained by using the ANN described in \cite{9921698}. 
Even if the MAE provided by ANN is lower than non-ML-based approaches (i.e., $2.04\%$), this approach is much more complex than EMA (whose MAE is $2.16\%$).
Therefore, ANNs are recommended only for specific application contexts where the device executing the model has enough power in terms of computation resources.

The cumulative distribution function (CDF) reported in Fig~\ref{fig:CDF} shows the empirical probability that any given method provides an error less than or equal to a given threshold (reported in abscissa). 
Starting from the CDF, percentiles can be easily derived as the value on the x-axis that corresponds to a given percentage on the y-axis (for instance, $95\%$ for $e_{\mathrm{p}_{95}}$). 
As expected, this set of curves confirms that EMA outperforms the other models practically always.

\begin{table*}[t]
  \caption{Lower bounds on errors, obtained when an Oracle is used to select the best among the other models (OR model).}
  \label{tab:res_oracle}
  \small
  \begin{center}
    \tabcolsep=0.22cm
    \def\arraystretch{1.25}
    \begin{tabular}{c|c|ccccccc}
    Model & $\mathrm{MSE}$ & $\mathrm{MAE}$ & $\sigma_e$ & $e_{\mathrm{p}_{90}}$ & $e_{\mathrm{p}_{95}}$ & $e_{\mathrm{p}_{99}}$ & $e_{\mathrm{p}_{99.9}}$ & $e_{\mathrm{max}}$ \\
    & $[\cdot 10^{-3}]$ & \multicolumn{7}{c}{[\%]} \\
    \hline
    OR (ALL) & 0.397 & 1.27 & 1.54 & 3.31 & 4.16 & 7.06 & 12.40 & 13.89 \\
    \hline
    OR (EMA+WMA) & 0.762 & 2.05 & 1.85 & 4.48 & 5.52 & 8.40 & 13.53 & 14.86 \\
    OR (EMA+SMA) & 0.740 & 1.98 & 1.86 & 4.40 & 5.44 & 8.52 & 13.53 & 14.86 \\
    OR (EMA+SLR) & 0.589 & 1.77 & 1.67 & 3.89 & 4.76 & 7.39 & 13.28 & 13.89 \\
    OR (EMA+PR2)  & 0.710 & 1.89 & 1.88 & 4.33 & 5.45 & 8.49 & 13.53 & 14.86 \\
    OR (EMA+PR3)  & 0.647 & 1.83 & 1.77 & 4.14 & 5.29 & 8.24 & 12.45 & 14.30 \\
    \hline
    OR (EMA+SLR+PR3) & 0.452 & 1.47 & 1.54 & 3.44 & 4.29 & 7.15 & 12.40 & 13.89 \\
    \hline
    \end{tabular}
    \end{center}
    \vspace{-0.25cm}
\end{table*}

\begin{table*}[t]
  \caption{Statistics on the prediction error achieved by the EMA model for different values of $\alpha$.}
  \label{tab:EMA}
  \small
  \begin{center}
    \tabcolsep=0.22cm
    \def\arraystretch{1.25}
    \begin{tabular}{c|c|c|ccccccc|c}
    Model & $\alpha$ & $\mathrm{MSE}$ & $\mathrm{MAE}$ & $\sigma_e$ & $e_{\mathrm{p}_{90}}$ & $e_{\mathrm{p}_{95}}$ & $e_{\mathrm{p}_{99}}$ & $e_{\mathrm{p}_{99.9}}$ & $e_{\mathrm{max}}$ & $w$ \\
    & & $[\cdot 10^{-3}]$ & \multicolumn{7}{c|}{[\%]} & [\%] \\
    \hline
    EMA & 0.000125 & 0.810 & 2.24 & 1.75 & 4.73 & 5.43 & 6.92 & 11.80 & 12.18 & 39.22 \\
    EMA & 0.000375 & 0.803 & 2.16 & 1.83 & 4.52 & 5.58 & 8.52 & 13.53 & 14.86 & 22.89 \\
    EMA & 0.001125 & 0.907 & 2.28 & 1.97 & 4.72 & 5.89 & 9.71 & 14.18 & 16.16 & 37.89 \\
    \hline
    COM (EMA) & - & 0.733 & 2.10 & 1.71 & 4.22 & 5.29 & 8.07 & 12.82 & 14.38 & - \\    
    \hline    OR (EMA) & - &  0.400 & 1.38 & 1.44 & 3.21 & 4.28 & 6.29 & 11.51 & 12.18 & - \\ 
    \hline
    \end{tabular}
    \end{center}
    \vspace{-0.3cm}
\end{table*}

The best technique, when the number of times a model wins over the others is considered, is SLR, which wins $27.58\%$ of the times. 
Unfortunately, when it does not win a non-negligible prediction error is often experienced, which makes the average result in terms of MAE worse than the other models. 
Curiously, EMA is the best model by looking at the statistical indicators, but at the same time it is the one that more seldom wins, i.e., $w=7.54\%$. 
This probably depends on the fact that, even when losing, 
EMA manages to provide satisfactory estimates.

\subsection{Oracle}
Starting from the results about the winning percentage, the performance of the oracle (OR) model described in Subsection~\ref{sub:oracle} was analyzed, with the aim to provide a meaningful lower bound on statistical indicators. 
OR selects on a sample-by-sample basis the model that experiences the lowest error.
Results about the oracle are reported in Table~\ref{tab:res_oracle}. 
In the first row, identified by ``OR (ALL)'', the oracle performs its selection among all the six models analyzed in this work by using (\ref{eq:OR}). 
Behavior is quite interesting, because OR achieves an absolute reduction of the MAE by $0.89\%$ (from $2.16\%$ to $1.27\%$) if compared to the best EMA model. 
As can be seen from the values in the table, all other statistical indexes were also consistently improved when resorting to the oracle.

These results open an interesting research direction about the use of machine learning, or artificial intelligence in general, to predict channel quality.
In particular, an ANN can be exploited for adaptively selecting the best prediction model in order to try approaching the accuracy of the OR model,
instead of using it to directly foresee an estimate for the FDR in the future.
In this respect, the OR model can be seen as a lower bound, which can be hardly achieved in practice but that might be approached by an algorithm that takes care of the automatic selection of the best model. 
This investigation is out of the scope of this work, but we deem it promising.
As such, it will be the subject of our future research activities.

In the following rows of Table~\ref{tab:res_global}, the OR model was applied to select between the (best) EMA model and one of the other five models. 
Quite interesting, it can be noted that selecting between EMA and SLR achieves a MAE equal to $1.77\%$. 
This means that exploiting just two models, which are not unreasonably complex to implement, potentially enables a $0.39\%$ reduction of the MAE (from $2.16\%$ to $1.77\%$).

Running OR with three models, i.e., EMA, SLR, and PR3, yields higher benefits, with a $0.69\%$ reduction of MAE (from $2.16\%$ to $1.47\%$). 
We found that using more than three models is not convenient, because the error does not shrink tangibly but, at the same time, complexity of the machine learning system grows, as the number of options that need to be classified increases.

\subsection{Considerations about EMA}
In this subsection, the model which provided the best results, i.e., EMA, is analyzed for different values of $\alpha$. 
In addition to the optimal value $\alpha^*=0.000375$, which minimizes the MSE over the whole database $\mathcal{D}_{tr}$, the results obtained with $\alpha=\alpha^*/3=0.000125$ and $\alpha=3 \cdot \alpha^*=0.001125$ are also reported in Table~\ref{tab:EMA}. 

As expected, both are worse than the case when $\alpha=\alpha^*$.
However, analyzing the number of times when an EMA model with a specific value of $\alpha$ wins (column $w$), it can be seen that the model with $\alpha=0.000125$ has a $39.22\%$ chance of winning, while the model with $\alpha=0.001125$ wins $37.89\%$ of the times. 
This means that, sometimes, it is better to have a more reactive model (higher values of $\alpha$) to quickly track variations of the channel quality, whereas when channel conditions are stable selecting a lower value for $\alpha$ makes the EMA model more accurate.

Again, an oracle (``OR (EMA)'' model) can be used to find a reasonable lower bound on the prediction error, by selecting the value of $\alpha$ on a frame-by-frame basis.
The results we obtained are quite promising, with the MSE that lowers from $2.16\%$ (EMA model with $\alpha=\alpha^*$) down to $1.38\%$ for the OR model. 
In the OR case, also all the other statistical indices about accuracy are significantly better.
This suggests that a machine learning algorithm can be used not only to select the best model among a set of options, but as a classifier to dynamically select the best parameterization, i.e., the optimal $\alpha$ in the case of EMA or the optimal $N_p$ for the other models. 
Again, this is left for future work.

Last but not least, the linear combination of the three EMA models with equal weights ($\lambda^{(m)}=\frac{1}{3}$), denoted ``COM (EMA)'', achieved a MAE equal to $2.10\%$, which closely resembles what we obtained with the ANN model.

\section{Conclusions}
\label{sec:conclusions}
In this work, the ability to predict the future quality of a wireless (Wi-Fi) channel in terms of the frame delivery ratio, 
starting from the outcomes of past transmission attempts, was investigated using a number of simple models based on moving averages and regression.

The exponential moving average, typically known as EMA, showed the lowest prediction error when compared to the other models. EMA enables lightweight (inexpensive) implementations in terms of computational and memory requirements, which makes it a perfect choice for devices with limited resources, like motes in wireless sensor (and actuator) networks or, more in general, embedded devices.

Performing a linear combination among the estimates produced by multiple EMA models permits to lower the prediction error further. 
By doing so, a mean absolute error as low as $2.10\%$ was obtained in our experiments, which is comparable to what can be achieved using much more complex models like those based on artificial neural networks.

This research work paves the way to the definition of advanced techniques aimed at improving prediction accuracy further, which either rely on suitable linear combinations of outcomes generated by other models or that make use of machine learning algorithms to dynamically select the best prediction model depending on the observed channel conditions.
In our future work transmission latency will be additionally considered.

\bibliographystyle{IEEEtran}
\bibliography{bibliography}

\begin{thebibliography}{10}
\providecommand{\url}[1]{#1}
\csname url@samestyle\endcsname
\providecommand{\newblock}{\relax}
\providecommand{\bibinfo}[2]{#2}
\providecommand{\BIBentrySTDinterwordspacing}{\spaceskip=0pt\relax}
\providecommand{\BIBentryALTinterwordstretchfactor}{4}
\providecommand{\BIBentryALTinterwordspacing}{\spaceskip=\fontdimen2\font plus
\BIBentryALTinterwordstretchfactor\fontdimen3\font minus
  \fontdimen4\font\relax}
\providecommand{\BIBforeignlanguage}[2]{{%
\expandafter\ifx\csname l@#1\endcsname\relax
\typeout{** WARNING: IEEEtran.bst: No hyphenation pattern has been}%
\typeout{** loaded for the language `#1'. Using the pattern for}%
\typeout{** the default language instead.}%
\else
\language=\csname l@#1\endcsname
\fi
#2}}
\providecommand{\BIBdecl}{\relax}
\BIBdecl

\bibitem{9714203}
S.~Sudhakaran, K.~Montgomery, M.~Kashef, D.~Cavalcanti, and R.~Candell,
  ``{Wireless Time Sensitive Networking Impact on an Industrial Collaborative
  Robotic Workcell},'' \emph{IEEE Transactions on Industrial Informatics},
  vol.~18, no.~10, pp. 7351--7360, 2022.

\bibitem{CANAS2021107379}
H.~Cañas, J.~Mula, M.~Díaz-Madroñero, and F.~Campuzano-Bolarín,
  ``{Implementing Industry 4.0 principles},'' \emph{Computers \& Industrial
  Engineering}, vol. 158, p. 107379, 2021.

\bibitem{MADDIKUNTA2022100257}
\BIBentryALTinterwordspacing
P.~K.~R. Maddikunta, Q.-V. Pham, P.~B, N.~Deepa, K.~Dev, T.~R. Gadekallu,
  R.~Ruby, and M.~Liyanage, ``{Industry 5.0: A survey on enabling technologies
  and potential applications},'' \emph{Journal of Industrial Information
  Integration}, vol.~26, p. 100257, 2022. [Online]. Available:
  \url{https://www.sciencedirect.com/science/article/pii/S2452414X21000558}
\BIBentrySTDinterwordspacing

\bibitem{8401919}
E.~Sisinni, A.~Saifullah, S.~Han, U.~Jennehag, and M.~Gidlund, ``{Industrial
  Internet of Things: Challenges, Opportunities, and Directions},'' \emph{IEEE
  Transactions on Industrial Informatics}, vol.~14, no.~11, pp. 4724--4734,
  2018.

\bibitem{9779183}
M.~Friesen, L.~Wisniewski, and J.~Jasperneite, ``{Machine Learning for
  Zero-Touch Management in Heterogeneous Industrial Networks - A Review},'' in
  \emph{IEEE 18th International Conference on Factory Communication Systems
  (WFCS 2022)}, 2022, pp. 1--8.

\bibitem{SCANZIO2021103388}
S.~Scanzio, L.~Wisniewski, and P.~Gaj, ``{Heterogeneous and dependable networks
  in industry – A survey},'' \emph{Computers in Industry}, vol. 125, p.
  103388, 2021.

\bibitem{9945847}
J.~Perez-Ramirez, O.~Seijo, and I.~Val, ``{Time-Critical IoT Applications
  Enabled by Wi-Fi 6 and Beyond},'' \emph{IEEE Internet of Things Magazine},
  vol.~5, no.~3, pp. 44--49, 2022.

\bibitem{electronics11030304}
G.~Cena, S.~Scanzio, and A.~Valenzano, ``{Ultra-Low Power Wireless Sensor
  Networks Based on Time Slotted Channel Hopping with Probabilistic
  Blacklisting},'' \emph{Electronics}, vol.~11, no.~3, 2022.

\bibitem{LEONARDI202257}
L.~Leonardi, L.~{Lo Bello}, and G.~Patti, ``{LoRa support for long-range
  real-time inter-cluster communications over Bluetooth Low Energy industrial
  networks},'' \emph{Computer Communications}, vol. 192, pp. 57--65, 2022.

\bibitem{2016-TII-WiRed}
G.~{Cena}, S.~{Scanzio}, and A.~{Valenzano}, ``{Seamless Link-Level Redundancy
  to Improve Reliability of Industrial Wi-Fi Networks},'' \emph{IEEE Trans.
  Ind. Informat.}, vol.~12, no.~2, pp. 608--620, April 2016.

\bibitem{9921559}
J.~Haxhibeqiri, P.~A. Campos, I.~Moerman, and J.~Hoebeke, ``{Safety-related
  Applications over Wireless Time-Sensitive Networks},'' in \emph{IEEE 27th
  International Conference on Emerging Technologies and Factory Automation
  (ETFA 2022)}, 2022, pp. 1--8.

\bibitem{9903301}
S.~Scanzio, G.~Cena, and A.~Valenzano, ``{Enhanced Energy-Saving Mechanisms in
  TSCH Networks for the IIoT: The PRIL Approach},'' \emph{IEEE Transactions on
  Industrial Informatics}, pp. 1--11, 2022.

\bibitem{5502548}
T.~Watteyne, S.~Lanzisera, A.~Mehta, and K.~S.~J. Pister, ``{Mitigating
  Multipath Fading through Channel Hopping in Wireless Sensor Networks},'' in
  \emph{2010 IEEE International Conference on Communications}, 2010, pp. 1--5.

\bibitem{705532}
M.~Simon and M.~Alouini, ``{A unified approach to the performance analysis of
  digital communication over generalized fading channels},'' \emph{Proceedings
  of the IEEE}, vol.~86, no.~9, pp. 1860--1877, 1998.

\bibitem{9786784}
S.~Szott, K.~Kosek-Szott, P.~Gawłowicz, J.~T. Gómez, B.~Bellalta, A.~Zubow,
  and F.~Dressler, ``{Wi-Fi Meets ML: A Survey on Improving IEEE 802.11
  Performance With Machine Learning},'' \emph{IEEE Communications Surveys \&
  Tutorials}, vol.~24, no.~3, pp. 1843--1893, 2022.

\bibitem{9120030}
A.~K. Gizzini, M.~Chafii, A.~Nimr, and G.~Fettweis, ``{Deep Learning Based
  Channel Estimation Schemes for IEEE 802.11p Standard},'' \emph{IEEE Access},
  vol.~8, pp. 113\,751--113\,765, 2020.

\bibitem{8884240}
A.~Kulkarni, A.~Seetharam, A.~Ramesh, and J.~D. Herath, ``{DeepChannel:
  Wireless Channel Quality Prediction Using Deep Learning},'' \emph{IEEE Trans.
  Veh. Technol.}, vol.~69, no.~1, pp. 443--456, 2020.

\bibitem{8813020}
W.~Jiang and H.~D. Schotten, ``{Neural Network-Based Fading Channel Prediction:
  A Comprehensive Overview},'' \emph{IEEE Access}, vol.~7, pp.
  118\,112--118\,124, 2019.

\bibitem{9781119562306}
\BIBentryALTinterwordspacing
W.~Jiang, H.~Dieter~Schotten, and J.-y. Xiang, \emph{{Neural Network–Based
  Wireless Channel Prediction}}.\hskip 1em plus 0.5em minus 0.4em\relax John
  Wiley \& Sons, Ltd, 2020, ch.~16, pp. 303--325. [Online]. Available:
  \url{https://onlinelibrary.wiley.com/doi/abs/10.1002/9781119562306.ch16}
\BIBentrySTDinterwordspacing

\bibitem{2022-ITL-ML}
\BIBentryALTinterwordspacing
S.~Scanzio, F.~Xia, G.~Cena, and A.~Valenzano, ``{Predicting Wi-Fi link quality
  through artificial neural networks},'' \emph{Internet Technology Letters},
  vol.~5, no.~2, p. e326, 2022. [Online]. Available:
  \url{https://onlinelibrary.wiley.com/doi/abs/10.1002/itl2.326}
\BIBentrySTDinterwordspacing

\bibitem{9921698}
S.~Scanzio, G.~Cena, C.~Zunino, and A.~Valenzano, ``{Machine Learning to
  Support Self-Configuration of Industrial Systems Interconnected over
  Wi-Fi},'' in \emph{IEEE 27th International Conference on Emerging
  Technologies and Factory Automation (ETFA 2022)}, 2022, pp. 1--8.

\bibitem{MONGELLI20161}
M.~Mongelli and S.~Scanzio, ``{A neural approach to synchronization in wireless
  networks with heterogeneous sources of noise},'' \emph{Ad Hoc Networks},
  vol.~49, pp. 1--16, 2016.

\bibitem{6817598}
------, ``{Approximating Optimal Estimation of Time Offset Synchronization With
  Temperature Variations},'' \emph{IEEE Transactions on Instrumentation and
  Measurement}, vol.~63, no.~12, pp. 2872--2881, 2014.

\bibitem{4769479}
W.~T. Zaw and T.~T. Naing, ``{Modeling of Rainfall Prediction over Myanmar
  Using Polynomial Regression},'' in \emph{2009 International Conference on
  Computer Engineering and Technology}, vol.~1, 2009, pp. 316--320.

\bibitem{8747047}
T.~Bakibayev and A.~Kulzhanova, ``Common movement prediction using polynomial
  regression,'' in \emph{{IEEE 12th International Conference on Application of
  Information and Communication Technologies (AICT 2018)}}, 2018, pp. 1--4.

\bibitem{7991945}
G.~Cena, S.~Scanzio, and A.~Valenzano, ``{A software-defined MAC architecture
  for Wi-Fi operating in user space on conventional PCs},'' in \emph{2017 IEEE
  13th International Workshop on Factory Communication Systems (WFCS)}, 2017,
  pp. 1--10.

\bibitem{8477080}
------, ``{SDMAC: A Software-Defined MAC for Wi-Fi to Ease Implementation of
  Soft Real-Time Applications},'' \emph{IEEE Transactions on Industrial
  Informatics}, vol.~15, no.~6, pp. 3143--3154, 2019.

\end{thebibliography}

\end{document}